\documentclass[11pt]{article}
\usepackage{moriond,epsfig}

\def\PRL{{\em Phys. Rev. Lett.}}
\def\PRD{{\em Phys. Rev.} D}

\def\mco{\multicolumn}

\def\be{\begin{equation}}
\def\ee{\end{equation}}
\def\bea{\begin{eqnarray}}
\def\eea{\end{eqnarray}}

\def\D0{D\O}                            
\def\met{\mbox{${\hbox{$E$\kern-0.6em\lower-.1ex\hbox{/}}}_T$}} 
\begin{document}
\vspace*{4cm}
\title{DIBOSON CROSS SECTIONS AT $\sqrt{s}$=1.96 TeV}

\author{A. W. ASKEW \\
(for the CDF and \D0 Collaborations)
}

\address{Fermilab, MS 352, \D0 Experiment, P.O. Box 500,\\
Batavia, IL 60510, USA}

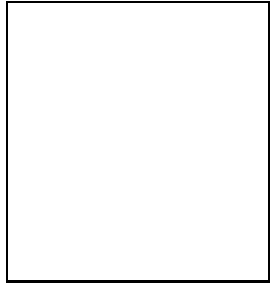
\maketitle\abstracts{
A brief survey of the results on diboson production at the
Tevatron is presented.  Measured cross sections for $W\gamma$, $Z\gamma$, 
$WW$, and limits on WZ/ZZ are summarized.}

\section{Introduction}

Diboson production gives information about the structure of the 
Standard Model (SM) electroweak interaction.
By studying the production of vector boson pairs,
one may determine whether the interaction is behaving as the
SU(2)$_L\otimes$U(1)$_Y$ gauge symmetry in the SM, or 
whether the structure of the theory is entirely different.
There is no mechanism which alters the vector boson couplings within the SM. 
Any deviation is a sign of new physics.  In all cases the presence of anomalous 
couplings manifests itself in an increase in the cross section for pairs of vector bosons.
Thus measurement of the diboson production cross sections is a test of the theory. 

In these proceedings, a brief summary of the current results from
the Tevatron Run II is presented.  Where possible, the main kinematic
and fiducial requirements imposed by the CDF and \D0 experiments are
detailed, and references to the more detailed publications provided~\footnote{A
detailed description of the two Tevatron experiments is beyond the scope of
these proceedings, however the interested reader is referred to
~\cite{RunIICDFDet} and ~\cite{RunIID0Det}.}.

\section{W$\gamma$}
  
Anomalous coupling at the WW$\gamma$ vertex leads to not only a higher cross section, 
but also an excess of high transverse energy ($E_T$) photons.  The $W\gamma$ 
cross section is measured with respect to a lepton-photon separation requirement and a 
threshold on the $E_T$ of the photon.  This avoids a divergences in the theoretical
calculation where the photon is colinear with the lepton, and where the photon is very 
low $E_T$.  CDF and \D0 both use a lepton-photon separation cut of 
$\Delta {\cal R} > 0.7$.  CDF and \D0 use different photon $E_T$ thresholds, 
so the results are not directly comparable.  The dominant background for both 
experiments is W+jet production where the jet mimics a photon.


The kinematic and fiducial requirements for the CDF and \D0 $W\gamma$ analyses 
are summarized in Table~\ref{tab:ALLWgamma}.  CDF measures the cross section 
$\sigma(p\overline{p} \rightarrow W\gamma + X)$ 
with $E_T^\gamma>7$ GeV and $\Delta {R}_{\ell\gamma}>0.7$ is 
to be 18.3~$\pm$~3.1~pb (stat. and syst. combined), in good agreement 
with the SM value of 19.3~$\pm$~1.4~pb.  This analysis has 
been published~\cite{CDFWGammaZGamma}.
\begin{table}[htbp]
\caption{Summary of CDF and \D0 Fiducial and Kinematic Requirements for $W\gamma$.
Estimated SM and background events are included, along with observed candidates.\label{tab:ALLWgamma}}
\vspace{0.4cm}
\begin{center}
\begin{tabular}{|c|c|c|c|c|}
\hline
\mco{1}{|c|}{} & \mco{2}{|c|}{CDF} & \mco{2}{|c|}{\D0}  \\
\hline
Channel: 		& $e\nu\gamma$ 	& $\mu\nu\gamma$ & $e\nu\gamma$ 	& $\mu\nu\gamma$  \\
\hline
$\eta^{\ell}$		& 2.6		& 1.0		& 1.1 			& 2.0 \\
$p_{T}^{\ell}$		& 25		& 20		& 25  			& 20  \\
$\met$			& 25		& 20		& 25  			& 20  \\
$\eta^{\gamma}$		& 1.0		& 1.0		& 1.1 			& 1.1 \\  
$p_{T}^{\gamma}$	& 7		& 7		& 8   			& 8   \\
$M_{T}$			&30$<M_T<$120	& 30$<M_T<$120	& $M_T >$ 40		& 0   \\
\hline
Lumi. ($pb^{-1}$)	& 202		& 192		& 162			& 134 \\
Background		& 67.3$\pm$18.1 & 47.3$\pm$7.6	& 60.8$\pm$ 4.5 	& 71.3$\pm$ 5.2\\
Expected (SM)		& 126.8$\pm$5.8	& 95.2$\pm$4.9	& 59.5$\pm$ 5.4		&  94.0$\pm$7.4\\
\hline
Candidates		& 195		& 128		& 112			& 161 \\
\hline
\end{tabular}
\end{center}
\end{table}
\D0 measures the cross section 
$\sigma(p\overline{p} \rightarrow W\gamma + X)$
with $E_T^\gamma>8$ GeV and $\Delta {\cal R}_{\ell\gamma}>0.7$ is to be 14.8~$\pm$2.1~pb (stat. and syst. combined), in 
good agreement with the SM value of 16.0~$\pm$~0.4 pb. 

\D0 uses the $E_{T}^{\gamma}$ distribution to set limits on anomalous 
vector boson couplings.  The one-dimensional limits on the anomalous coupling parameters
are:   $-0.88 < \Delta\kappa_\gamma < 0.96$ and $-0.20 < \lambda_\gamma < 0.20$.  
This is a preliminary result from \D0~\cite{D0WGamma}.


\section{Z$\gamma$}

In the SM, the Z and the photon do not directly couple to each other
at leading order.    
Anomalous couplings would manifest as an excess of events at high photon $E_T$
and an increase in the cross section.
As with $W\gamma$, the $Z\gamma$ cross section is measured 
with respect to a lepton-photon separation requirement and a threshold on the 
$E_T$ of the photon.  Both experiments have made cuts consistent with
their $W\gamma$ analyses for the photon $E_T$ threshold, and the lepton
photon separation requirement.  The only significant background is Z+jet production 
where the jet mimics a photon.


The selection criteria for the CDF and \D0 $Z\gamma$ analyses are summarized in Table~\ref{tab:ALLZgamma}.
CDF measures the cross section
$\sigma(p\overline{p} \rightarrow Z\gamma + X)$
with $E_T^\gamma>7$ GeV and $\Delta {\cal R}_{\ell\gamma}>0.7$ to be 4.6~$\pm$0.6~pb, in
good agreement with the SM value of 4.5$\pm$0.3~pb.  This analysis has been published
in~\cite{CDFWGammaZGamma}.
\begin{table}[htbp]
\caption{Summary of CDF and \D0 Fiducial and Kinematic Requirements for $Z\gamma$.
Estimated SM and background events are included, along with observed candidates.\label{tab:ALLZgamma}}
\vspace{0.4cm}
\begin{center}
\begin{tabular}{|c|c|c|c|c|}
\hline
\mco{1}{|c|}{} & \mco{2}{|c|}{CDF} & \mco{2}{|c|}{\D0}  \\
\hline
Channel: 		& $ee\gamma$ 	& $\mu\mu\gamma$ & $ee\gamma$ 	& $\mu\mu\gamma$  \\
\hline
$\eta^{\ell}$		& 2.6		& 1.0		 & 1.1 (2.5) 	& 2.0 \\
$p_{T}^{\ell}$		& 25		& 20		 & 25		& 15  \\
$\eta^{\gamma}$		& 1.0		& 1.0		 & 1.1		& 1.1 \\  
$p_{T}^{\gamma}$	& 7		& 7		 & 8		& 8 \\
$M_{\ell\ell}$		&40$<M_{\ell\ell}<$130	& 40$<M_{\ell\ell}<$130	& 30$<M_{\ell\ell}$ & 30$<M_{\ell\ell}$ \\
\hline
Lumi. ($pb^{-1}$)	& 202		& 192		& 320		& 290 \\
Background		& 2.8$\pm$0.9   & 2.1$\pm$0.6	& 23.6$\pm$2.3	& 22.4$\pm$3.0\\
Expected (SM)		& 31.3$\pm$1.6	& 33.6$\pm$1.5	& 95.3$\pm$4.9	& 126.0$\pm$7.8\\
\hline
Candidates		& 36		& 35		& 138		& 152\\
\hline
\end{tabular}
\end{center}
\end{table}
\D0 measures the cross section
$\sigma(p\overline{p} \rightarrow Z\gamma + X$
with $E_T^\gamma>8$ GeV and $\Delta {\cal R}_{\ell\gamma}>0.7$ to be 4.2~$\pm$0.5~pb, in
good agreement with the SM value of 3.9$\pm$0.2~pb.

\D0 uses the photon $E_T$ spectrum to set limits on anomalous couplings.
The 95\% C.L. limits are 
$|h_{10,30}^{Z}|<0.23$, $|h_{20,40}^{Z}|<0.020$, $|h_{10,30}^{\gamma}|<0.23$, and 
$|h_{20,40}^{\gamma}|<0.019$ (for a form factor scale of $\Lambda=1~TeV$.  The limits on
$|h_{20,40}^{Z}|$ and $|h_{20,40}^{\gamma}|$ represent the most stringent limits on these
couplings to date.  This analysis has been submitted for publication~\cite{D0Zgamma}.


\section{WW}
The WW final state has couplings to both the photon and the Z.  Anomalous coupings in W-pair
production are heavily constrained by studies performed at LEP.  However, WW is a favored decay
channel for the Higgs boson, and additional production at high center of
mass energy could give evidence for other non-SM heavy resonances.  

WW production has several physics backgrounds from other lepton pair production.  Drell-Yan production,
WZ production, ZZ production, and even top quark pairs all have channels in which at least two
leptons can be produced, and thus must be addressed for the purpose of identifying true WW events.
In both analyses presented here, cuts have optimized to reduce these backgrounds.

\subsection{CDF WW Results}

Similar cuts to those described in the previous analyses for identifying leptons from W decays
are used to identify W-pairs.  Leptons (e, $\mu$) are required to be identified within the
fiducial coverage ($|\eta|<2.0$ for electrons,$|\eta|<1.0$ for muons).  The leptons are required
to be $p_T>$ 20 GeV/$c$.  The event $\met$ is required to be greater than 25 GeV.  
The significance of the $\met$ is required to be greater than 3, and
a veto is imposed on events that have jets above 15 GeV within $|\eta| <$2.5, to reduce 
contamination from top quark events.  CDF identifies a total of 17 candidates (6 in the ee, 
5 in the e$\mu$ and 6 in the $\mu\mu$ channels respectively), and using a luminosity
of 184~pb$^{-1}$ measures a cross section
of $14.6^{+5.8}_{-5.1}(stat.)^{+1.8}_{-3.0}(syst.)\pm$~0.6~pb, in good agreement with the
SM value of 12.4~$\pm$~0.8~pb.  This analysis has been submitted for publication~\cite{CDFWW}.

\subsection{\D0 WW Results}

Similar cuts to those described in the previous analyses for identifying leptons from W decays
are used to identify W-pairs.  Leptons (e, $\mu$) are required to be identified within the
fiducial coverage ($|\eta|<3.0$ for electrons,$|\eta|<2.0$ for muons).  The leptons are required
to be at $p_T >$20~GeV/$c$ for the lead lepton and $p_T > $15 GeV/$c$ for the trailing lepton.  
The event $\met$ is required to be greater than 30 GeV in the di-electron channel, 20 GeV in the
e-$\mu$ channel, and 40 GeV in the di-muon channel.
The scaled $\met$ is required to be greater than 15 in the
di-electron and e-$\mu$ channel.  A cut is imposed on the sum of the $E_T$ of jets 
within $|\eta| < 2.5$ and $E_T >$20, for the di-electron and e-$\mu$ channels at 50 GeV, and for
the di-muon channel at 100 GeV to limit contamination from top quark events.  
\D0 identifies a total of 25 candidates (6 in the ee, 
15 in the e$\mu$ and 4 in the $\mu\mu$ channels), and using a luminosity of
approximately 230~pb$^{-1}$ measures a cross section
of $13.8^{+4.3}_{-3.8}(stat.)^{+1.2}_{-0.9}(syst.)\pm$~0.9~pb, in good agreement with the
SM value of 12.4~$\pm$~0.8~pb.  This analysis has been accepted for publication~\cite{D0WW}.


\section{CDF WZ/ZZ Results}

CDF combines a search for WZ and ZZ together in a number of different topologies, and sets a 
cross section limit.
Two leptons ($ee$, $\mu\mu$) are required to resolve the invariant mass for the Z boson
(required to be within the range 76~GeV$ <M_{\ell\ell}<$106~GeV).
Then three separate event topologies were considered:  two leptons plus $\met$, three leptons
plus $\met$ and four leptons.  The only topology with candidate events was two leptons plus
$\met$.  Three candidate events(two di-electron, one dimuon) remain 
after all selection cuts.  Using a luminosity of 184~pb$^{-1}$ CDF proceeded to set a cross section limit 
of $\sigma(p\overline{p}\rightarrow ZW/ZZ) < 15.2~$pb.  This analysis has been submitted
for publication~\cite{CDFZWZZ}.

\section{\D0 WZ Results}

\D0 performed a search for WZ events in the three leptons plus $\met$ channel exclusively.
Two of the leptons were required to reconstruct to the Z mass range (for di-eletrons 
 $71~GeV <M_{\ell\ell}<111~GeV$ and for di-muons  51~GeV$ <M_{\ell\ell}<$131~GeV).  
Three candidates were observed (two tri-muon, and 1 tri-electron).
Using approximately 290~pb$^{-1}$ of luminosity \D0 sets a cross 
section limit of $\sigma(p\overline{p}\rightarrow ZW/ZZ) < 13.3~$pb.  Using these three
candidates, \D0 sets limits on anomalous WZ couplings.  The one-dimensional limits are:
$-0.53 < \lambda_{Z} < 0.56$, $-0.57 < \Delta g_{1}^{Z} < 0.76$, and $-2.0 < \Delta\kappa_{Z} < 2.4$.
This is a preliminary result from \D0~\cite{D0WZ}.

\end{document}